\documentclass[11pt,twoside]{article}
\usepackage{asp2014}

\aspSuppressVolSlug
\resetcounters
\begin{document}

\title{IACT event analysis with the MAGIC telescopes using deep convolutional neural networks with CTLearn}

\author{T. Miener$^1$, R. L\'{o}pez-Coto$^2$, J. L. Contreras$^1$, J. G. Green$^3$, D. Green$^4$ for the MAGIC Collaboration, E. Mariotti$^2$, D. Nieto$^1$, L. Romanato$^2$, and S. Yadav$^5$}
\affil{$^1$Instituto de F\'{i}sica de Part\'{i}culas y del Cosmos and Departamento de EMFTEL, Universidad Complutense de Madrid, Spain; \email{tmiener@ucm.es}}
\affil{$^2$ Dipartimento di Fisica e Astronomia dell?Università and Sezione INFN, Padova,Italy}
\affil{$^3$INAF - National Institute for Astrophysics, Roma, Italy}
\affil{$^4$Max-Planck-Institut für Physik, München, Germany} 
\affil{$^5$Birla Institute of Technology and Science, Pilani, India} 

\paperauthor{T. Miener}{tmiener@ucm.es}{https://orcid.org/0000-0003-1821-7964}{Universidad Complutense de Madrid, Madrid, Spain}{Instituto de F\'{i}sica de Part\'{i}culas y del Cosmos and Departamento de EMFTEL}{Madrid}{Madrid}{E-28040}{Spain}
\paperauthor{R. L\'{o}pez-Coto}{ruben.lopezcoto@pd.infn.it}{https://orcid.org/0000-0002-3882-9477}{Padova}{Physics Department}{Padova}{Italy}{00000}{Italy}
\paperauthor{E. Mariotti}{mr.ektor@gmail.com}{}{Padova}{Physics Department}{Padova}{Italy}{00000}{Italy}
\paperauthor{J. L. Contreras}{jlcontreras@fis.ucm.es}{https://orcid.org/0000-0001-7282-2394}{Universidad Complutense de Madrid, Madrid, Spain}{Instituto de F\'{i}sica de Part\'{i}culas y del Cosmos and Departamento de EMFTEL}{Madrid}{Madrid}{E-28040}{Spain}
\paperauthor{D. Green}{damgreen@mpp.mpg.de}{https://orcid.org/0000-0003-0768-2203
}{MPI Munich}{Physics Department}{Munich}{Bavaria}{00000}{Germany}
\paperauthor{J. Green}{jarred.green@inaf.it}{https://orcid.org/0000-0002-1130-6692
}{INAF}{Physics Department}{Italy}{Italy}{00000}{Italy}
\paperauthor{D. Nieto}{d.nieto@ucm.es}{https://orcid.org/0000-0003-3343-0755}{Universidad Complutense de Madrid, Madrid, Spain}{Instituto de F\'{i}sica de Part\'{i}culas y del Cosmos and Departamento de EMFTEL}{Madrid}{Madrid}{E-28040}{Spain}

\begin{abstract}
\small
The Major Atmospheric Gamma Imaging Cherenkov (MAGIC) telescope system consists of two imaging atmospheric Cherenkov telescopes (IACTs) and is located on the Canary island of La Palma. IACTs are excellent tools to inspect the very-high-energy (few tens of GeV and above) gamma-ray sky by capturing images of the air showers, originated by the absorption of gamma rays and cosmic rays by the atmosphere, through the detection of Cherenkov photons emitted in the shower. One of the main factors determining the sensitivity of IACTs to gamma-ray sources, in general, is how well reconstructed the properties (type, energy, and incoming direction) of the primary particle triggering the air shower are. We present how deep convolutional neural networks (CNNs) are being explored as a promising method for IACT full-event reconstruction. The performance of the method is evaluated on observational data using the standard MAGIC Analysis and Reconstruction Software, \texttt{MARS}, and \texttt{CTLearn}, a package for IACT event reconstruction through deep learning.

\end{abstract}

\section{Introduction}
\small
In this contribution, we show how deep learning (DL) algorithms like CNNs are incorporated into the analysis workflow of the MAGIC telescopes to perform full-event reconstruction. We also explore the robustness of CNN-based methods, when applying them to real observational data and compare the sensitivity to the standard analysis of \texttt{MARS}~\citep{Zanin:2013,Aleksic:2016}. The DL workflow consists of three main building bricks (see Fig.~\ref{fig:MAGICDL_workflow}). The Monte Carlo (MC) simulations and observational data are processed by the \texttt{MARS} software. A complementary macro translate crucial information into \texttt{uproot}\footnote{\href{https://github.com/scikit-hep/uproot4}{https://github.com/scikit-hep/uproot4}}-readable branches~\citep{pivarski:2021}. Then, the \texttt{DL1-Data-Handler}\footnote{\href{https://github.com/cta-observatory/dl1-data-handler}{https://github.com/cta-observatory/dl1-data-handler}}~\citep{kim:2020} (DL1DH) assembles several data levels from the standard approach and unifies them in a common data format in \texttt{HDF5} designed for DL purposes. The training of the CNN-based models and their prediction, the actual full-event reconstruction, are performed with \texttt{CTLearn}\footnote{\href{https://github.com/ctlearn-project/ctlearn}{https://github.com/ctlearn-project/ctlearn}}~\citep{Nieto:2019ak,brill:2019}, a backend for IACT analysis using \texttt{TensorFlow}.

\articlefigure[width=1.0\textwidth]{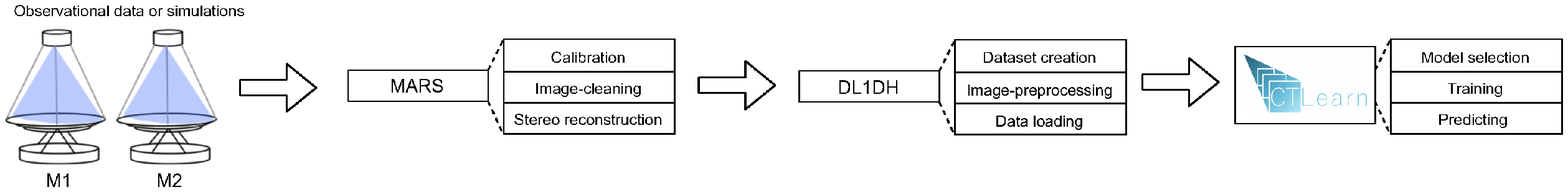}{fig:MAGICDL_workflow}{Diagram depicting the main analysis steps of the MAGIC DL analysis with CTLearn.}

\section{DL analysis with the MAGIC telescopes}
\small
\paragraph{Model selection} For this work, \texttt{CTLearn}’s Thin-ResNet (TRN) was selected based on previous studies ~\citep{Grespan:2021,Miener:2021}. Stereoscopic information are explored by concatenating the images (integrated pixel charges and signal arrival times) of MAGIC1 and MAGIC2 channel-wise before feeding the network. We explored two different analysis schemes, where we trained the same TRN model with raw images, containing besides the Cherekov light of the shower also the fluctuations of the Night Sky Background (NSB), and cleaned images, where pixels dominated by noise rather than Cherenkov light emission are set to zero. The cleaning mask are obtained with the standard \texttt{MARS} analysis cleaning. Since the pixel layout of the MAGIC cameras is a hexagonal lattice, we mapped them to a Cartesian lattice using bilinear interpolation to directly apply CNNs~\citep{Nieto:2019uj}.

\paragraph{Validation on simulations} To evaluate the performance common metrics like ROC curves, energy and angular resolution curves are used, applying the same quality cuts (see Fig.~\ref{fig:Simulation_validation}). The reconstruction performance is obtained using MC gamma simulations coming uniformly from a $ 0.4^{\circ} $ offset of the telescope pointing (ringwobble). To demonstrate the robustness of CNNs trained with cleaned images, we tested all methods for the background rejection against MC proton simulations and observational off data, where we do not expect any gamma-ray signal.

\articlefigurethree{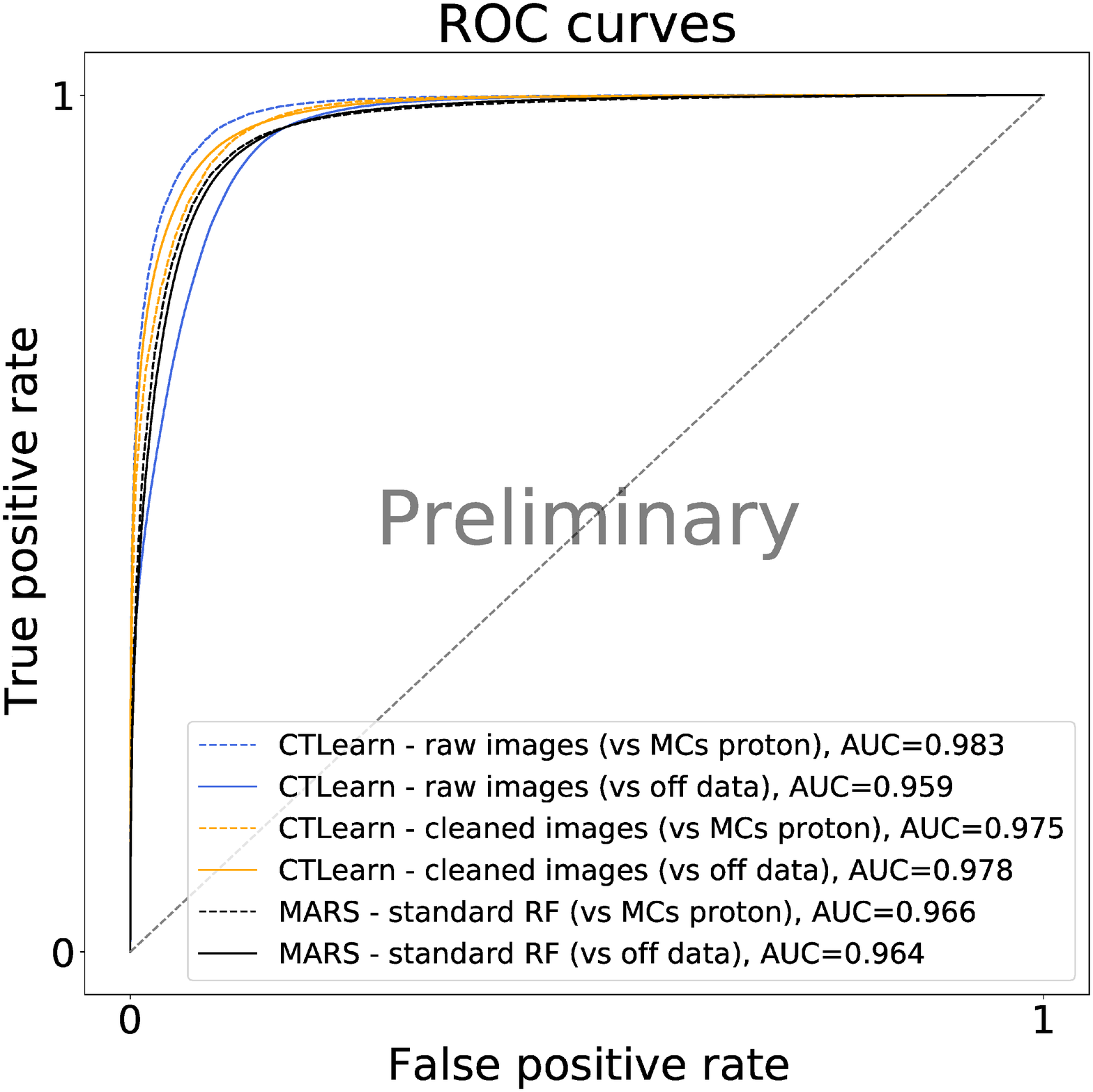}{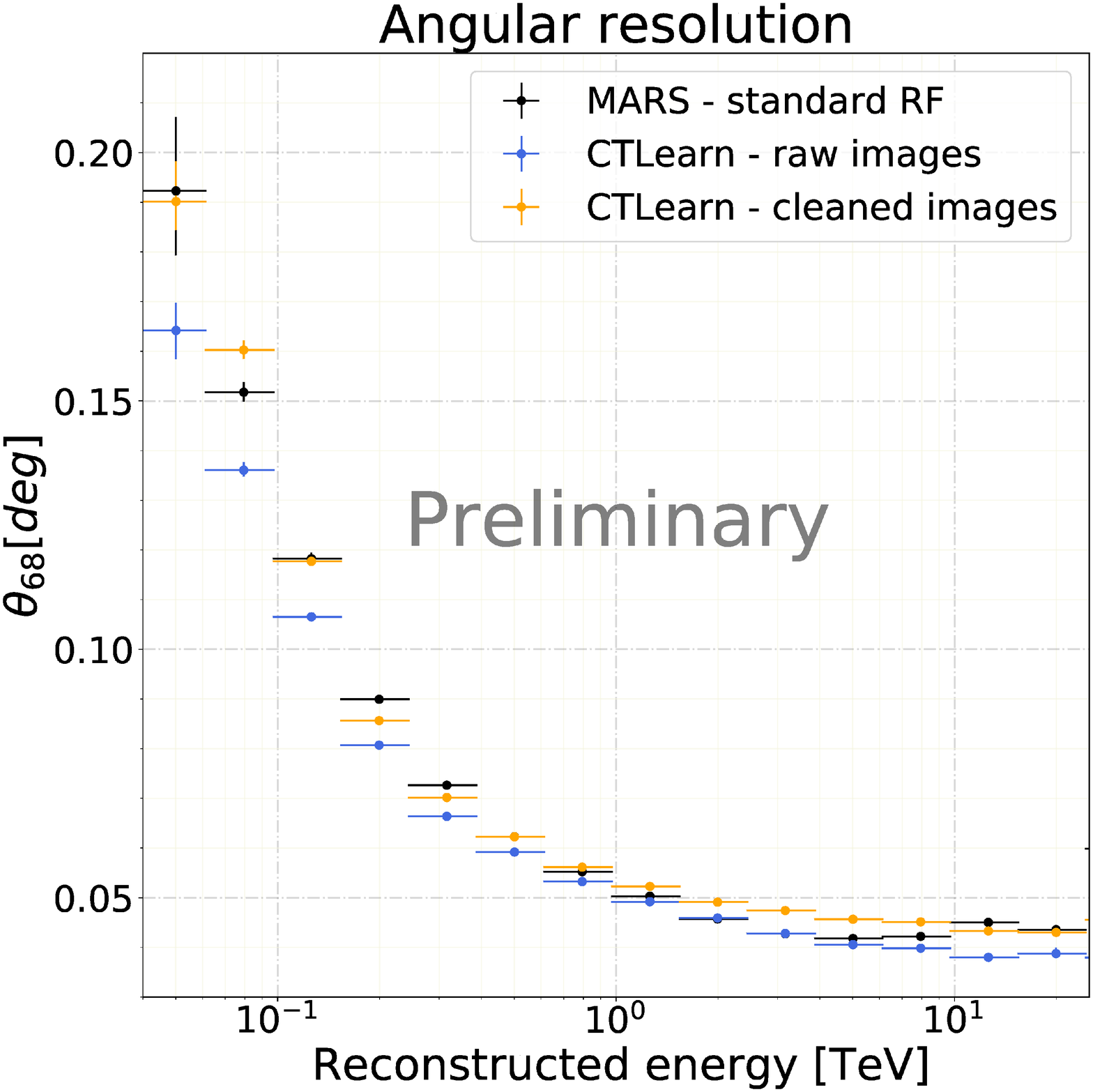}{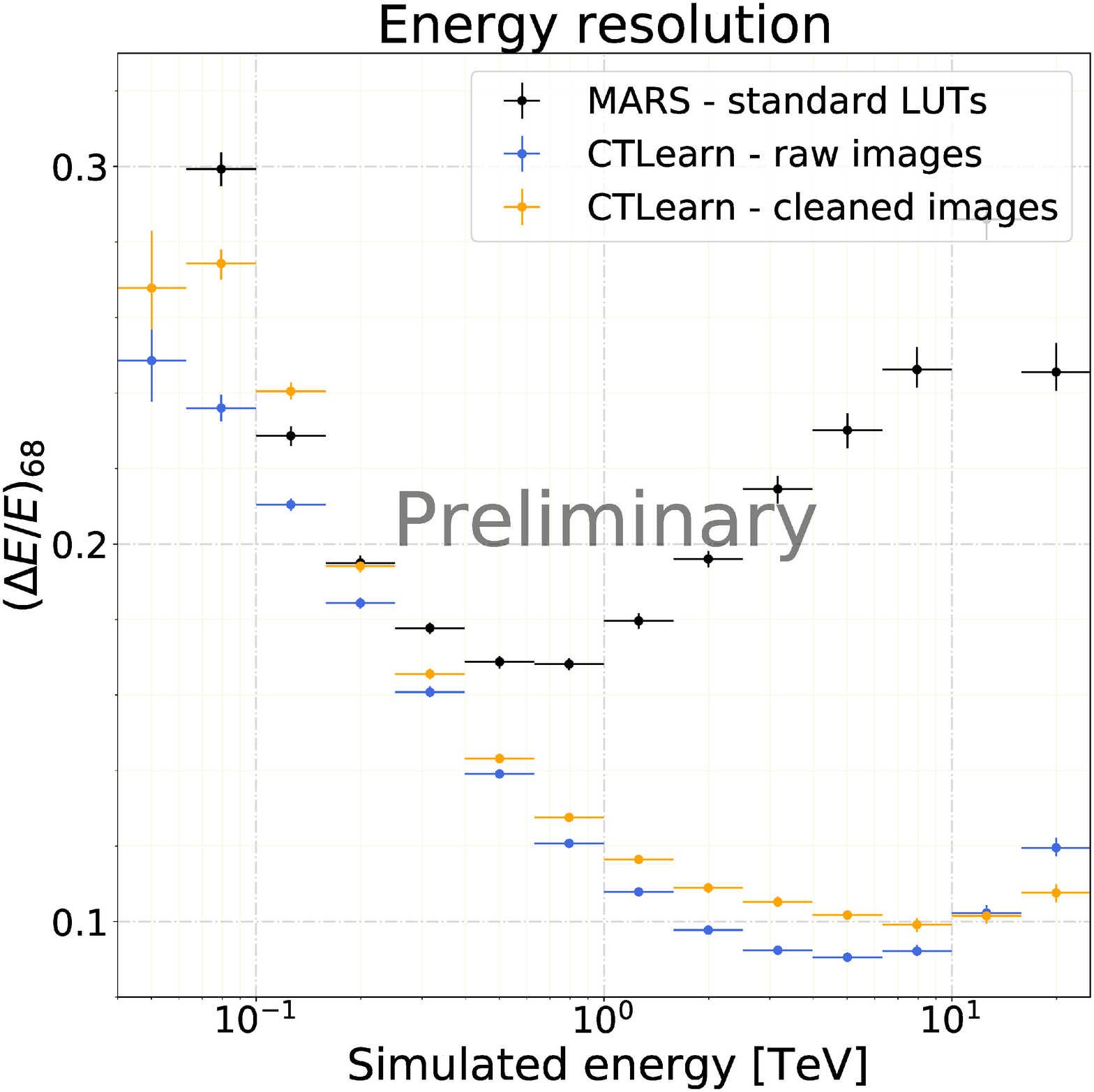}{fig:Simulation_validation}{The performance parameters are obtained using the MC gamma simulations (ringwobble). \emph{Left)} ROC curves tested against MC proton simulations and observational off data. \emph{Center)} Angular resolution vs. reconstructed energy. \emph{Right)} Energy resolution vs. simulated energy.}

\paragraph{Results on observational data}

In this work, 2.93h observation of the standard gamma-ray candle Crab Nebula, taken on four different nights in 2018 under good weather condition at low zenith (zd < $35^{\circ}$), are considered. The data have been analyzed with latest \texttt{MARS} software using the standard settings for the analysis focusing of the medium energy (ME) - and low energy (LE) range. For \texttt{CTLearn}, we strictly adopted the quality cuts from the \texttt{MARS} analysis. ME analysis (> $250$ GeV) apply the cuts: valid stereo reconstruction, $\theta^{2}$ < 0.009 $\text{deg}^{2}$, hadronness < 0.16 and both hillas intensity sizes > 300 phe, while the LE analysis (> $100$ GeV) apply the cuts: valid stereo reconstruction, $\theta^{2}$ < 0.02 $\text{deg}^2$, hadronness < 0.28 and both hillas intensity sizes > 60 phe. To fairly compare the results, obtained with CNN-based models, with the standard approach (random forest (RF) for the background rejection, Look-Up tables (LUTs) for the energy estimation and RF for bidimensional direction reconstruction), the hadronness cut is adjusted in the \texttt{CTLearn} analysis to equalize the background (bkg) rates for the corresponding standard MARS analyses (ME or LE). A source detection is determined using a $\theta^{2}$ plot (see Fig.~\ref{fig:theta2CTLearnME} for the \texttt{CTLearn} ME analysis with cleaned images) and the significance (Sig. in Tab.~\ref{tab:results_summary}) calculation (Eq. 17 in~\citep{LiMa:1983}). The main properties of all analyses are summarized in Tab.~\ref{tab:results_summary}. The sensitivity (Sen. in Tab.~\ref{tab:results_summary}) is computed as the strength of the source that gives excess/sqrt(background) = 5 after 50h.
   
\begin{table}  
\centering
\resizebox{1\textwidth}{!}{
    \begin{tabular}{|c|c|c|c|c|c|c|c|} 
        \hline
        Analysis & $ N_{on} $ & $ N_{off} $ & $ N_{ex} $ & $ \gamma $ rate [/min]& bkg rate [/min] & Sen. [\% Crab] & Sig. (Li\&Ma) \\
        \hline
        \hline
        MARS – ME & $ 819 $ & $21.0\pm2.6$ & $798.0\pm28.7$ & $4.54\pm0.16$ & $0.119\pm0.015$ & $0.70\pm0.05$ & $43.0\sigma$\\
        \hline
        CTLearn – ME (raw) & $ 629 $ & $23.3\pm3.1$ & $605.7\pm25.3$ & $3.45\pm0.14$ & $0.133\pm0.018$ & $0.97\pm0.08$ & $36.5\sigma$\\
        \hline
        CTLearn – ME (cleaned) & $ 844 $ & $22.0\pm2.7$ & $822.0\pm29.2$ & $4.68\pm0.17$ & $0.125\pm0.015$ & $0.69\pm0.05$ & $43.6\sigma$\\
        \hline
        \hline
        MARS – LE & $ 3579 $ & $679.0\pm15.0$ & $2900.0\pm61.7$ & $16.49\pm0.35$ & $3.861\pm0.086$ & $1.09\pm0.03$ & $61.1\sigma$\\
        \hline
        CTLearn – LE (raw) & $ 2730 $ & $673.7\pm20.0$ & $2056.3\pm56.0$ & $11.70\pm0.32$ & $3.832\pm0.114$ & $1.53\pm0.05$ & $47.5\sigma$\\
        \hline
        CTLearn – LE (cleaned) & $ 3536 $ & $680.7\pm15.1$ & $2855.3\pm61.3$ & $16.24\pm0.35$ & $3.872\pm0.086$ & $1.11\pm0.03$ & $60.4\sigma$\\
        \hline
     \end{tabular}}
    \caption{Summary of all performed analyses of the same Crab Nebula sample.}
    \label{tab:results_summary}
\end{table}

\articlefigure[width=0.95\textwidth]{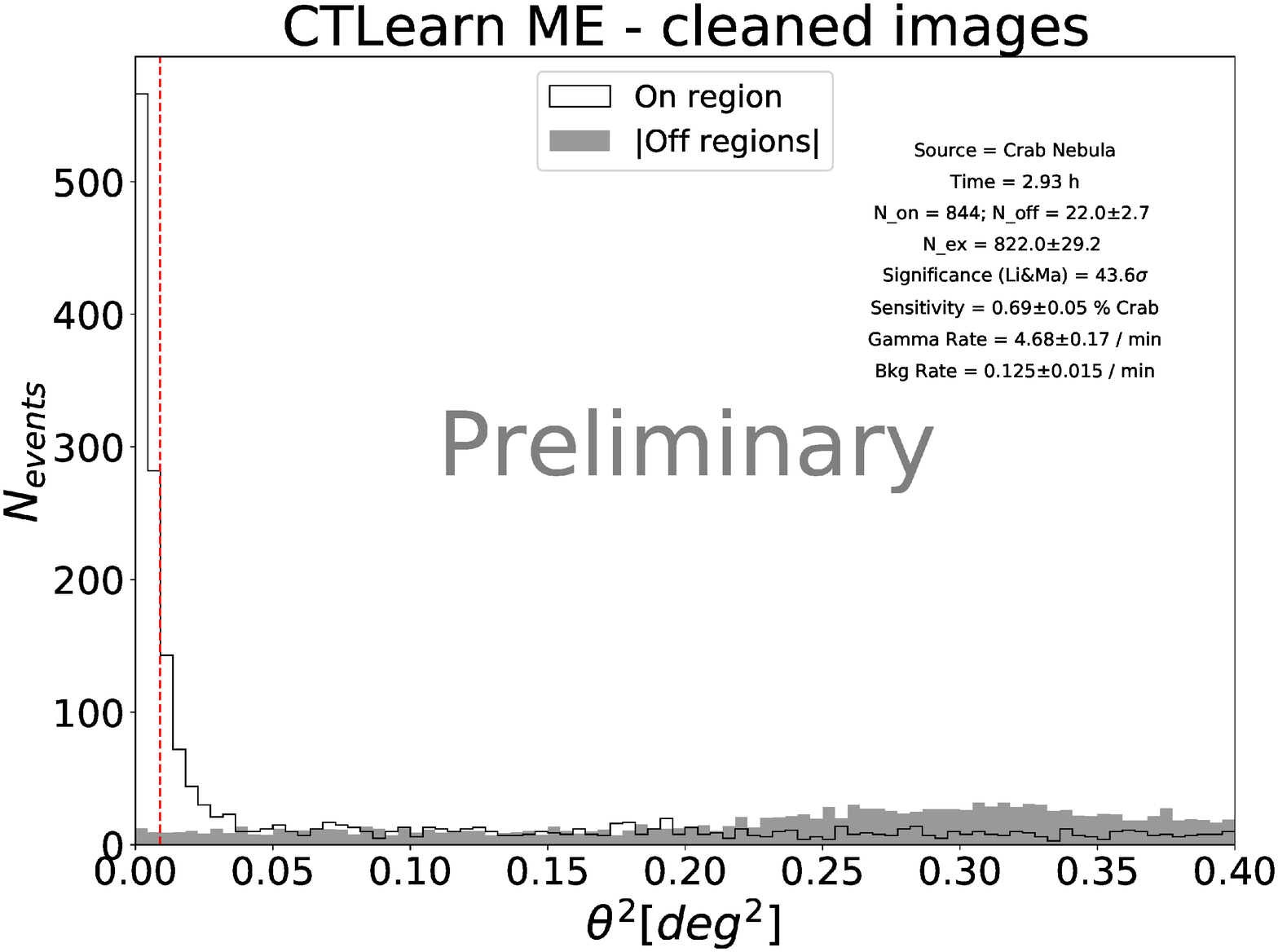}{fig:theta2CTLearnME}{$\theta^{2}$ plot for the CTLearn ME analysis with cleaned images.}

\section{Conclusions and outlook}
\small
This contribution shows for the first time that CNN-based full-event reconstruction works for the MAGIC telescopes and that \texttt{CTLearn} analyses are capable of detecting the Crab Nebula with a clear signal. We demonstrate that CNNs trained with cleaned images rather than raw images show a stronger robustness, when applying them to observational data, and the performance already matches the sensitivity of detection of the conventional analysis on real data. The performance of CNNs trained with raw images can be optimized by pixel-wise tuning of the NSB noise of the MCs~\citep{Vuillaume:2021} to match the NSB level of each observation. The selected TRN model is relatively shallow and further performance enhancements are foreseen by increasing the model depth/complexity. Future work is planned, where the full performance of CNNs under various observation conditions are evaluated.

\acknowledgements \scriptsize We would like to thank the Instituto de Astrofísica de Canarias for the excellent working conditions at the Observatorio del Roque de los Muchachos in La Palma. The financial support of the German BMBF, MPG and HGF; the Italian INFN and INAF; the Swiss National Fund SNF; the ERDF under the Spanish Ministerio de Ciencia e Innovación (MICINN) (FPA2017-87859-P, FPA2017- 85668-P, FPA2017-82729-C6-5-R, FPA2017-90566-REDC, PID2019-104114RB-C31, PID2019-104114RB-C32, PID2019- 105510GB-C31C42, PID2019-~107847RB-C44, PID2019-107988GB-C22); the Indian Department of Atomic Energy; the Japanese ICRR, the University of Tokyo, JSPS, and MEXT; the Bulgarian Ministry of Education and Science, National RI Roadmap Project DO1-268/16.12.2019 and the Academy of Finland grant nr. 317637 and 320045 are gratefully acknowledged. This work was also supported by the Spanish Centro de Excelencia “Severo Ochoa” SEV-2016- 0588, SEV-2017-0709 and CEX2019-000920-S, and "María de Maeztu” CEX2019-000918-M, the Unidad de Excelencia “María de Maeztu” MDM-2015-0509-18-2 and the "la Caixa" Foundation (fellowship LCF/BQ/PI18/11630012), by the Croatian Science Foundation (HrZZ) Project IP-2016-06-9782 and the University of Rĳeka Project 13.12.1.3.02, by the DFG Collaborative Research Centers SFB823/C4 and SFB876/C3, the Polish National Research Centre grant UMO-2016/22/M/ST9/00382 and by the Brazilian MCTIC, CNPq and FAPERJ.
TM acknowledges support from PID2019-104114RB-C32. JLC and DN acknowledges partial support from The European Science Cluster of Astronomy \& Particle Physics ESFRI Research Infrastructures funded by the European Union’s Horizon 2020 research and innovation program under Grant Agreement no. 824064. SY acknowledges financial support from Google LLC through the Google Summer of Code 2020 program. We acknowledge the support of NVIDIA Corporation with the donation of a Titan X Pascal GPU used for part of this research.
\\
\\
This paper has gone through internal review by the MAGIC Collaboration.

\tiny
\bibliography{X0-007}

\end{document}